\documentclass[a4paper]{amsart}

\RequirePackage{amsmath}
\RequirePackage{bm}
\RequirePackage{amssymb}
\RequirePackage{upref}
\RequirePackage{amsthm}
\RequirePackage{enumerate}
\RequirePackage{pb-diagram}
\RequirePackage{amsfonts}
\RequirePackage[mathscr]{eucal}
\RequirePackage{verbatim}
\RequirePackage{xr}
\RequirePackage{graphicx}
\usepackage{calc}
\usepackage{xspace}
\RequirePackage{color}
\RequirePackage{ifthen}

\newcommand{\cf}{cf.\@\xspace}
\newcommand{\resp}{resp.\@\xspace}

\newcommand{\al}{\alpha}
\newcommand{\bet}{\beta}
\newcommand{\ga}{\gamma}
\newcommand{\de}{\delta }

\newcommand{\f}{\varphi}
\newcommand{\h}{\eta}

\newcommand{\lam}{\lambda}

\newcommand{\om}{\omega}

\newcommand{\C}{\varGamma}
\newcommand{\D}{\varDelta}
\newcommand{\F}{\varPhi}
\newcommand{\Lam}{\varLambda}
\newcommand{\Om}{\varOmega}

\newcommand{\fv}[2]{#1\hspace{0pt}_{|_{#2}}}

\newcommand{\so}{{\mc S_0}}

\newcommand{\msp[1]}[1]{\mspace{#1mu}}

\newcommand{\R}[1][n+1]{{\protect\mathbb R}^{#1}}

\newcommand{\Cc}{{\protect\mathbb C}}

\newcommand{\N}{{\protect\mathbb N}}

\newcommand{\eR}{\stackrel{\lower1ex \hbox{\rule{6.5pt}{0.5pt}}}{\msp[3]\R[]}}
\newcommand{\eN}{\stackrel{\lower1ex \hbox{\rule{6.5pt}{0.5pt}}}{\msp[1]\N}}
\newcommand{\eO}{\stackrel{\lower1ex \hbox{\rule{6pt}{0.5pt}}}{\msc O}}
\newcommand{\mf}[1]{\mathfrak {#1}}

\DeclareMathOperator{\supp}{supp}
\DeclareMathOperator{\id}{id}

\DeclareMathOperator{\rec}{Re}
\DeclareMathOperator{\imc}{Im}

\DeclareMathOperator{\Ad}{Ad}

\DeclareMathOperator{\diag}{diag}

\newcommand\im{\implies}
\newcommand\ra{\rightarrow}

\newcommand\hra{\hookrightarrow}

\newcommand\pa{\partial}
\newcommand\pde[2]{\frac {\partial#1}{\partial#2}}
 
\newcommand{\un}{\infty}
\newcommand{\A}{\forall}

\newcommand{\set}[2]{\{\,#1\colon #2\,\}}
\newcommand{\uu}{\cup}
\newcommand{\ii}{\cap}
\newcommand{\uuu}{\bigcup}

\newcommand{\uud}{ \stackrel{\lower 1ex \hbox {.}}{\uu}}
\newcommand{\uuud}[1]{ \stackrel{\lower 1ex \hbox {.}}{\uuu_{#1}}}
\newcommand\su{\subset}
\newcommand\Su{\Subset}

\newcommand\eS{\emptyset}
\newcommand{\sminus}[1][28]{\raise 0.#1ex\hbox{$\scriptstyle\setminus$}}

\newcommand{\wed}{\wedge}

\newcommand{\abs}[1]{\lvert#1\rvert}

\newcommand{\norm}[1]{\lVert#1\rVert}

\newcommand{\spd}[2]{\protect\langle #1,#2\protect\rangle}

\newcommand\cha[3]{{\bar\varGamma}_{#1#2}^#3}

\newcommand{\tbf}{\textbf}
\newcommand{\tit}{\textit}

\newcommand{\tup}{\textup}

\newcommand{\mc}{\protect\mathcal}
\newcommand{\msc}{\protect\mathscr}

\providecommand{\bysame}{\makebox[3em]{\hrulefill}\thinspace}

\newcommand{\bt}{\begin{thm}}
\newcommand{\bl}{\begin{lem}}
\newcommand{\bc}{\begin{cor}}
\newcommand{\bd}{\begin{definition}}
\newcommand{\bpp}{\begin{prop}}
\newcommand{\br}{\begin{rem}}
\newcommand{\bn}{\begin{note}}
\newcommand{\be}{\begin{ex}}
\newcommand{\bes}{\begin{exs}}
\newcommand{\bb}{\begin{example}}
\newcommand{\bbs}{\begin{examples}}
\newcommand{\ba}{\begin{axiom}}
\newcommand{\bas}{\begin{assumption}}

\newcommand{\et}{\end{thm}}
\newcommand{\el}{\end{lem}}
\newcommand{\ec}{\end{cor}}
\newcommand{\ed}{\end{definition}}
\newcommand{\epp}{\end{prop}}
\newcommand{\er}{\end{rem}}
\newcommand{\en}{\end{note}}
\newcommand{\ee}{\end{ex}}
\newcommand{\ees}{\end{exs}}
\newcommand{\eb}{\end{example}}
\newcommand{\ebs}{\end{examples}}
\newcommand{\ea}{\end{axiom}}
\newcommand{\eas}{\end{assumption}}

\newcommand{\bp}{\begin{proof}}
\newcommand{\ep}{\end{proof}}
\newcommand{\eps}{\renewcommand{\qed}{}\end{proof}}

\newcommand{\bal}{\begin{align}}

\newcommand{\bi}[1][1.]{\begin{enumerate}[\upshape #1]}
\newcommand{\bia}[1][(1)]{\begin{enumerate}[\upshape #1]}
\newcommand{\bin}[1][1]{\begin{enumerate}[\upshape\bfseries #1]}
\newcommand{\bir}[1][(i)]{\begin{enumerate}[\upshape #1]}
\newcommand{\bic}[1][(i)]{\begin{enumerate}[\upshape\hspace{2\cma}#1]}
\newcommand{\bis}[2][1.]{\begin{enumerate}[\upshape\hspace{#2\parindent}#1]}
\newcommand{\ei}{\end{enumerate}}

\newcommand\ndots{\raise 0.47ex \hbox {,}\hskip0.06em\cdots %
     \raise 0.47ex \hbox {,}\hskip0.06em} 

\newcommand{\q}{\quad}
\newcommand{\qq}{\qquad}

\newcommand{\hp}{\hphantom}

\newcommand\nd{\noindent}

\newskip\Csmallskipamount                                                
\Csmallskipamount=\smallskipamount
\newskip\Cmedskipamount
\Cmedskipamount=\medskipamount
\newskip\Cbigskipamount
\Cbigskipamount=\bigskipamount

\newcommand\cvs{\vspace\Csmallskipamount}   
\newcommand\cvm{\vspace\Cmedskipamount}

\newskip\csa
\csa=\smallskipamount

\newskip\cma
\cma=\medskipamount

\newskip\cba
\cba=\bigskipamount

\newdimen\spt
\newcommand\citem{\cvs\advance\itemno by
1{(\romannumeral\the\itemno})\hskip3pt}
\newcommand{\bitem}{\cvm\nd\advance\itemno by
1{\bf\the\itemno}\hspace{\cma}}

\newcount\itemno
\newcommand{\las}[1]{\label{S:#1}}

\newcommand{\lae}[1]{\label{E:#1}}
\newcommand{\lat}[1]{\label{T:#1}}
\newcommand{\lal}[1]{\label{L:#1}}

\newcommand{\lar}[1]{\label{R:#1}}

\newcommand{\rs}[1]{Section~\ref{S:#1}}

\newcommand{\rt}[1]{Theorem~\ref{T:#1}}
\newcommand{\rl}[1]{Lemma~\ref{L:#1}}

\newcommand{\rr}[1]{Remark~\ref{R:#1}}

\newcommand{\re}[1]{\eqref{E:#1}}
\newcommand{\frt}[1]{Theorem~\ref{T:#1} on page~\tup{\pageref{T:#1}}}
\newcommand{\frl}[1]{Lemma~\ref{L:#1} on page~\tup{\pageref{L:#1}}}
\newcommand{\frr}[1]{Remark~\ref{R:#1} on page~\tup{\pageref{R:#1}}}

\newcommand{\fre}[1]{\eqref{E:#1} on page~\tup{\pageref{E:#1}}}

\newskip\thmskip
\thmskip=\parindent

\newskip\hsk
\newenvironment{hinw}{\labelsep=0pt\begin{list}{}{\labelsep=0pt\itemindent=0pt\labelwidth=0pt\leftmargin=\parindent\rightmargin=0pt\partopsep=\cba}%
\item\it\nopagebreak\nopagebreak}%
{\end{list}}

\newcommand\bh{\begin{hinw}}
\newcommand{\eh}{\end{hinw}}

\newtheoremstyle{normal}
  {\cba}
  {\cba}
  {}
  {\thmskip}
  {\bfseries}
  {.}
  {\hsk}
  {}

\newtheoremstyle{abschnitt}
  {\cba}
  {\cba}
  {}
  {\thmskip}
  {\bfseries}
  {.}
  {\hsk}
  {}

\newtheoremstyle{italic}
  {\cba}
  {\cba}
  {\itshape}
  {\thmskip}
  {\bfseries}
  {.}
  {\hsk}
  {}

\newtheoremstyle{aufgaben}
  {\cba}
  {\cba}
  {}
  {}
  {\normalsize\bfseries}
  {.}
  {\hsk}
  {}

\newtheoremstyle{break}
  {\cba}
  {\cba}
  {\itshape}
  {}
  {\bfseries}
  {.}
  {\newline}
  {}

\swapnumbers
\theoremstyle{italic}
\newtheorem{thm}[subsection]{Theorem}
\newtheorem{lem}[subsection]{Lemma}
\newtheorem{prop}[subsection]{Proposition}
\newtheorem{cor}[subsection]{Corollary}

\theoremstyle{normal}
\newtheorem{rem}[subsection]{Remark}
\newtheorem{definition}[subsection]{Definition}
\newtheorem{example}[subsection]{Example}
\newtheorem{examples}[subsection]{Examples}
\newtheorem{ex}[subsection]{Exercise}
\newtheorem{note}[subsection]{}
\newtheorem{axiom}[subsection]{Axiom}
\newtheorem{assumption}[subsection]{Assumption}

\theoremstyle{aufgaben}
\newtheorem{exs}[subsection]{Exercises}

\swapnumbers

\numberwithin{equation}{section}
\numberwithin{figure}{section}

\newenvironment{textequation}[1][0.8]
{\begin{equation}
\begin{aligned}
\begin{minipage}{#1\linewidth}}
{\end{minipage}
\end{aligned}
\end{equation}
\ignorespacesafterend}

\newcommand{\btext}{\begin{textequation}}
\newcommand{\etext}{\end{textequation}}

\def\hinweis{\@startsection{subsection}{2}%
 \z@{0.7\linespacing\@plus 0.5\linespacing}{0.7\linespacing}%
{\normalfont\itshape\indent}}

\newcounter{hours}\newcounter{minutes}
\newcommand{\printtime}{%
\setcounter{hours}{\time/60}%
\setcounter{minutes}{\time-\value{hours}*60}%
\ifthenelse{\value{minutes}<10}{\thehours :0\theminutes}{\thehours:\theminutes}}

\usepackage[german,english]{babel}
\usepackage{graphicx}
\RequirePackage{amsmath}
\RequirePackage{bm}
\RequirePackage{amssymb}
\RequirePackage{upref}
\RequirePackage{amsthm}
\RequirePackage{enumerate}
\RequirePackage{pb-diagram}
\RequirePackage{amsfonts}
\RequirePackage[mathscr]{eucal}
\RequirePackage{verbatim}
\RequirePackage{xr}
\RequirePackage{graphicx}
\usepackage{calc}
\usepackage{xspace}
\usepackage{dsfont}

\makeatletter
\RequirePackage{color}
\newcommand{\ann}[1]{\renewcommand{\@makefnmark}{\mbox{$^{\color{red}{\@thefnmark}}$}}%
\footnote {#1}}
\makeatother

\RequirePackage{upref}
\RequirePackage{amsthm}
\RequirePackage{enumerate}
\usepackage[mathscr]{eucal}

\usepackage{xr-hyper}

\usepackage{calc}

\newlength{\oddsidemarginlength}
\newlength{\topmarginlength}

\newcounter{numberoflines}
\newcounter{tempcc}
\oddsidemargin=\oddsidemarginlength
\evensidemargin=\oddsidemargin
\topmargin=\topmarginlength
\usepackage[colorlinks=true,linkcolor=blue,citecolor=blue,urlcolor=blue]{hyperref}  

\begin{document}

\flushbottom


\title[A unified quantum theory II]{A unified quantum theory II: gravity interacting with  Yang-Mills and spinor fields}

\author{Claus Gerhardt}
\address{Ruprecht-Karls-Universit\"at, Institut f\"ur Angewandte Mathematik,
Im Neuenheimer Feld 294, 69120 Heidelberg, Germany}
\email{\href{mailto:gerhardt@math.uni-heidelberg.de}{gerhardt@math.uni-heidelberg.de}}
\urladdr{\href{http://www.math.uni-heidelberg.de/studinfo/gerhardt/}{http://www.math.uni-heidelberg.de/studinfo/gerhardt/}}

\thanks{The work has been supported by the DFG}

\date{\today}
%


\begin{abstract} 
We quantize the interaction of gravity with Yang-Mills and spinor fields, hence offering a quantum theory incorporating all four fundamental forces of nature. Using canonical quantization we obtain solutions of the Wheeler-DeWitt equation in a vector bundle and the method of second quantization leads to  a symplectic vector space $(V,\om)$ and a corresponding CCR representation for the bosonic components and a CAR representation for the fermionic part. The solution space of the Wheeler-DeWitt equation is invariant under  gauge transformations and under  isometries in the spacelike base space $\so$ of a given Riemannian metric $\rho_{ij}$. We also define a net of local subalgebras which satisfy four of the Haag-Kastler axioms.
\end{abstract}

\maketitle

\tableofcontents

\setcounter{section}{0}
\section{Introduction}
A unified quantum theory incorporating the four fundamental forces of nature is one of the major open problems in physics. The Standard Model combines electro-magnetism, the strong force and the weak force, but ignores gravity. The quantization of gravity is therefore a necessary first step to achieve a unified quantum theory.

The Einstein equations are the Euler-Lagrange equations of the Einstein-Hilbert functional and quantization of a Lagrangian theory requires to switch from a Lagrangian view to a Hamiltonian view. In a ground breaking paper, Arnowitt, Deser and Misner \cite{adm:old} expressed the Einstein-Hilbert Lagrangian in a form which allowed to derive a corresponding Hamilton function by applying the Legendre transformation. However, since the Einstein-Hilbert Lagrangian is singular, the Hamiltonian description of gravity is only correct if two additional constraints are satisfied, namely, the Hamilton constraint and the diffeomorphism constraint. Dirac \cite{dirac:lqm} proved how to quantize a constrained Hamiltonian system---at least in principle---and his method has been applied to the Hamiltonian setting of gravity, \cf the paper of DeWitt \cite{dewitt:gravity} and the monographs by Kiefer \cite{kiefer:book} and Thiemann \cite{thiemann:book}.  In the general case, when arbitrary globally hyperbolic spacetime metrics are allowed, the problem turned out to be extremely difficult and solutions could only be found by assuming a high degree of symmetry.

However, we recently achieved the quantization of gravity for general hyperbolic spacetimes, \cf \cite{cg:qgravity}, and, in a subsequent paper \cite{cg:uqtheory}, we developed a unified quantum theory for the interaction of gravity with a Yang-Mills field.

Using these results we are able to treat the interaction of gravity with Yang-Mills and spinor fields thereby offering a unified quantum theory for all four fundamental forces. Though we only consider the interaction of gravity with one Yang-Mills and one spinor field the inclusion of additional independent fields poses no problem.

We look at  the Lagrangian functional
\begin{equation}\lae{1.1}
\begin{aligned}
J&=\al_M^{-1}\int_{\tilde\Om}(\bar R-2\Lam)-\int_{\tilde\Om}\tfrac14 \ga_{\bar a\bar b}\bar g^{\mu\rho_2}\bar g^{\lam\rho_1}F^{\bar a}_{\mu\rho_1}F^{\bar b}_{\rho_2\lam}\\
&\hp =\;\;\,-\int_{\tilde\Om}\{\tfrac12 \bar g^{\mu\lam}\ga_{\bar a\bar b}\F^{\bar a}_\mu\bar\F^{\bar b}_\lam+V(\F)\}\\
&\hp =\;\;+\int_{\tilde\Om}\{\tfrac12[\tilde\psi_{I}E^\mu_a\ga^a(D_\mu \psi)^{I}+\overline{\tilde\psi_{I}E^\mu_a\ga^a(D_\mu \psi)^{I}}]+m\tilde\psi_{I}\psi^{I}\},
\end{aligned}
\end{equation}
where $\al_M$ is a positive coupling constant,  $\tilde\Om\Su N=N^{n+1}$ and $N$ a globally hyperbolic spacetime with metric $\bar g_{\al\bet}$, $0\le \al,\bet\le  n$.

As we proved in \cite{cg:qgravity} we may only consider metrics $\bar g_{\al\bet}$ that split with respect to some fixed globally defined time function $x^0$ such that
\begin{equation}\lae{1.2}
d\bar s^2=-w^2 (dx^0)^2+g_{ij}dx^idx^j
\end{equation}
where $g_{ij}(x^0,\cdot)$ are Riemannian metrics in $\so$,
\begin{equation}
\so=\{x^0=0\}.
\end{equation}

The functional $J$ consists of the Einstein-Hilbert functional, the Yang-Mills and Higgs functional and a massive Dirac term. Instead of the potential
\begin{equation}
m\tilde\psi_{I}\psi^{I}
\end{equation}
we could have considered an arbitrary potential as long as it is quadratic in $\psi$ and real.

The Yang-Mills field $(A_\mu)$
\begin{equation}
A_\mu=f_{\bar c}A^{\bar c}_\mu
\end{equation}
corresponds to the adjoint representation of a compact, semi-simple Lie group $\mc G$ with Lie algebra $\mf g$. The $f_{\bar c}$,
\begin{equation}
f_{\bar c}=(f^{\bar a}_{\bar c\bar b})
\end{equation}
are the structural constants of $\mf g$.

We assume the Higgs field $\F=(\F^{\bar a})$ to have complex valued components.

The spinor field $\psi=(\psi^I_A)$ has  a spinor index $A$, $1\le A\le n_1$, and a colour index $I$, $1\le I\le n_2$. Here, we suppose that the Lie group has a unitary representation $R$ such that
\begin{equation}
t_{\bar c}=R(f_{\bar c})
\end{equation}
are antihermitian matrices acting on $\Cc^{n_2}$. The symbol $A_\mu\psi$ is now defined by
\begin{equation}
A_\mu\psi=t_{\bar c}\psi A^{\bar c}_\mu
\end{equation}

In our previous papers we used canonical quantization to obtain a first quantization leading to the Wheeler-DeWitt equation
\begin{equation}
Hu=0,
\end{equation}
where the Hamiltonian $H$ is a symmetric, normally hyperbolic differential operator in a bundle $E$ with base space $\so$ and fibers which were considered to be globally hyperbolic Lorentzian manifolds equipped with a Lorentzian metric which was composed of the DeWitt metric and further Riemannian metrics resulting from the presence of the Yang-Mills and Higgs fields.

In a second step we had to use the method of second quantization to develop a quantum field theory for the solutions of the Wheeler-DeWitt equation leading to a real symplectic vector space $(V,\om)$ and a corresponding CCR representation. The CCR representation could be defined by a quantum field $\F_M$, where $M\su E$ was a Cauchy hypersurface.

In the present paper we follow this approach. First, we prove in \rs{3} that the Dirac Lagrangian $L_D$ (without the integration density) can be expressed in the form 
\begin{equation}
\begin{aligned}
L_D&=\tfrac{i}2(\bar\chi_I\dot\chi^I-\dot{\bar\chi}^{I}\chi_{I})w^{-1}\f^{-1}+m i\bar\chi_I\ga^0\chi^I\f^{-1}\\
&\q-\tfrac{i}2\{\bar\chi_I\ga^0E^k_{a'}\ga^{a'}\tilde D_k\chi^I-\overline{\bar\chi_I\ga^0E^k_{a'}\ga^{a'}\tilde D_k\chi^I}\}\f^{-1}.
\end{aligned}
\end{equation}
Here, we fixed a Riemann metric $\rho_{ij}$ on $\so$ and defined the function $\f$ on $N$ by
\begin{equation}\lae{1.11}
\f^2=\frac{\det(g_{ij})}{\det(\rho_{ij})},
\end{equation}
where the spacetime metric has the form as in \re{1.2}.

The spinor field $\chi=(\chi^I_A)$ is defined by
\begin{equation}
\chi =\sqrt\f\psi.
\end{equation}

Applying Casalbuoni's results in \cite{casalbuoni:odd} and \cite{casalbuoni:fermi} we obtain a corresponding Hamilton function 
\begin{equation}
\begin{aligned}
H_D&=
\tfrac{i}2\{\bar\chi_I\ga^0E^k_{a'}\ga^{a'}\tilde D_k\chi^I-\overline{\bar\chi_I\ga^0E^k_{a'}\ga^{a'}\tilde D_k\chi^I}\}\\
&\q-m i\bar\chi_I\ga^0\chi^I,
\end{aligned}
\end{equation} 
where $\bar \chi^A_I$ and $\chi^J_B$ satisfy the anticommutation rules
\begin{equation}
\{\bar\chi_{I}^A,\chi^{J}_B\}^*_+=-i\de^{J}_I\de_{B}^A.
\end{equation}

Canonical quantization---with $\bar h=1$---then requires that the corresponding operators $\hat{\chi}^{I}_A, \,\hat{\bar\chi}_{J}^B$ satisfy the anticommutative rules 
\begin{equation}
[\hat\chi^{I}_A,\hat{\bar\chi}_{J}^B]_+=i\{\chi^{I}_A,\bar\chi_{J}^B\}^*_+=\de^{I}_J\de_{A}^B
\end{equation}
and
\begin{equation}
[\hat{\bar\chi}_{I}^A,\hat{\bar\chi}_{J}^B]_+=[\hat\chi^{I}_A,\hat{\chi}^{J}_B]_+=0.
\end{equation}
We realize these quantum rules in the Grassmann algebra
\begin{equation}
\mc P=\mc P(\chi^I_A)
\end{equation}
consisting of polynomial functions
\begin{equation}
u=a_0+\sum_{k,m}a^{A_1\cdots A_m}_{I_1\cdots I_k}\chi^{I_1\cdots I_k}_{A_1\cdots A_m},
\end{equation}
where the coefficients are complex numbers being asymmetric in their indices, by defining
\begin{equation}
\bar\chi^A_I=\frac{\pa}{\pa\chi^I_A}
\end{equation}
to be the left derivative. The space $\mc P$ is also endowed with a natural Hermitian scalar product.

After quantization the Hamilton operator $H_D$ can be viewed as a self-adjoint operator in the finite dimensional Hilbert space $\mc P$.

We then define two bundles. The first bundle $E$ is essentially the bundle we already used in \cite{cg:uqtheory} with base space $\so$ and fibers
\begin{equation}
\mc F(x)=F(x)\times(\mf g\otimes T^{0,1}_x(\so))\times\mf g\times\mf g.
\end{equation}
The second bundle is a vector bundle $\hat E$ with base space $E$ and fiber
\begin{equation}
\hat{\mc  F}=T^{0,2}_{x,\tup{ symm}}(\so)\times(\mf g\otimes T^{0,1}_x(\so))\times\mf g\times \mc P.
\end{equation}
 Writing the elements $u$ of the fiber in coordinates we obtain
\begin{equation}
u=(f_{ij},A^{\bar a}_m,\F^{\bar b},w),
\end{equation}
where $w\in\mc P$,  the components $\F^{\bar b}$ are complex, while $f_{ij}$ and $A^{\bar a}_m$ are real. We consider $\hat{\mc  F}$ to be a real vector space and shall later equip it with a natural  real scalar product.

The common Hamilton function has the form 
\begin{equation}
\begin{aligned}
H&=H_G+H_{YM}+H_H+H_D\\
&\equiv\tilde H+H_D,
\end{aligned}
\end{equation}
and after quantization it will be transformed to a symmetric, normally hyperbolic differential operator $H$ acting only in the fibers of $E$. $H$ looks like
\begin{equation}
H=-\D+c,
\end{equation}
where $H_D$ is part of the zero order term $c$ and $H$ is defined for sections $u\in C^\un_c(E,\hat E)$. The solution space of the Wheeler-DeWitt equation 
\begin{equation}
Hu=0
\end{equation}
is identical with
\begin{equation}
G(C^\un_c(E,\hat E)),
\end{equation}
where $G$ is the Green's operator.

In order to apply the method of second quantization we split the fiber in $\hat E$ into its bosonic and fermionic components. For the bosonic components we obtain a  CCR representation of a symplectic vector space $(V,\om)$ and each Cauchy hypersurface $M\su E$ defines a quantum field $\F_M$ such that
\begin{equation}
W([u])=e^{i\F_M(u)},\qq [u]\in V,
\end{equation}
is a Weyl system for $(V,\om)$, while we construct a CAR representation and a corresponding quantum field for the fermionic part.

These results are proved in \rs{5} and \rs{6}. In the last section we define local subalgebras and prove that they satisfy four Haag-Kastler axioms.  

Let us conclude this section with the important remark:
\br
The choice of the metric $\rho_{ij}$ in \re{1.11} is arbitrary, but the metric should have a rich group of isometries, since these isometries as well as the gauge transformations act as isometries in the fibers of $\hat E$ and they leave the solution space of the Wheeler-DeWitt equation invariant. They are also symplectic transformations in $(V,\om)$.
\er

\section{Definitions and notations}
Greek indices $\al$, $\bet$ range from $0$ to $n$, Latin $i,j,k$ from $1$ to $n$ and we stipulate $0\le a, b\le n$ but $1\le a', b'\le n$. Barred indices $\bar a$ refer to the Lie algebra $\mf g$, $1\le \bar a\le n_0=\dim\mf g$.

$\ga_{\bar a\bar b}$ is the Cartan-Killing metric.

The Dirac matrices are denoted by $\ga^a$ and they satisfy
\begin{equation}
\ga^a\ga^b+\ga^b\ga^a=2\h^{ab}I,
\end{equation}
where $\h_{ab}$ is the Minkowski metric with signature $(-,+,\ldots,+)$. $\ga^0$ is antihermitian and $\ga^{a'}$ Hermitian.

The indices $a,b$ are always raised or lowered with the help of the Minkowski metric, Greek indices with the help of the spacetime metric $\bar g_{\al\bet}$.

The $\ga^a$ act in
\begin{equation}
\Cc^{2^{\frac{n+1}2}},
\end{equation}
if $n$ is odd and in
\begin{equation}
\Cc^{2^\frac n2}\oplus \;\Cc^{2^\frac n2},
\end{equation}
if $n$ is even. In both cases we simply refer to these spaces as
\begin{equation}
\Cc^{n_1},
\end{equation}
i.e., the spinor index $A$ has range $1\le A\le n_1$.

The colour index $I$ has range $1\le I\le n_2$ and hence a spinor field $\psi^I_A$ has values in
\begin{equation}
\Cc^{n_1}\otimes \Cc^{n_2}.
\end{equation}

Finally, a Hermitian form $\spd\cdot\cdot$ is antihermitian in the first argument.

\section{Spinor fields}\las{3}
The Lagrangian of the spinor field is stated in \re{1.1}. Here, $\psi=(\psi^I_A)$ is a multiplet of spinors with spin $\tfrac12$; $A$ is the spinor index, $1\le A\le n_1$, and $I$, $1\le I\le n_2$, the \tit{colour} index. 
We shall also lower or raise the index $I$ with the help of the Euclidean metric $(\de_{IJ})$.

Let $\C_\mu$ be the spinor connection
\begin{equation}
\C_\mu=\tfrac14 \om_{\mu\hp ba}^{\hp\mu b}\ga_b\ga^a,
\end{equation}
then the covariant derivative $D_\mu\psi$ is defined by
\begin{equation}
D_\mu\psi=\psi_{,\mu}+\C_\mu\psi+A_\mu\psi.
\end{equation}

Let $(e^b_\lam)$ be a $n$-bein such that
\begin{equation}
\bar g_{\mu\lam}=\h_{ab}e^a_\mu e^b_\lam,
\end{equation}
where $(\h_{ab})$ is the Minkowski metric, and let $(E^\mu_a)$ be its inverse 
\begin{equation}
E^\mu_a=\h_{ab}\bar g^{\mu\lam}e^b_\lam,
\end{equation}
\cf \cite[p. 246]{eguchi:book}. 

The covariant derivative of $E^\al_a$ with respect to $(\bar g_{\al\bet})$ is then given by
\begin{equation}
E^\al_{a;\mu}=E^\al_{a,\mu}+\cha \mu\bet\al E^\bet_a
\end{equation}
and
\begin{equation}
\om_{\mu\hp ba}^{\hp\mu b}=E^\lam_{a;\mu}e^b_\lam=-E^\lam_ae^b_{\lam;\mu},
\end{equation}
hence the spin connection $\C_\mu$ can be expressed as
\begin{equation}
\C_\mu=\tfrac14 \om_{\mu\hp ba}^{\hp\mu b}\ga_b\ga^a=\tfrac14 E^\lam_{a;\mu}e^b_\lam \ga_b\ga^a=-\tfrac14 E^\lam_ae^b_{\lam;\mu}\ga_b\ga^a. 
\end{equation}

We shall first show:
\bl\lal{3.1}
Let $\bar g_{\al\bet}$ be a fixed spacetime metric that is split by the time function $x^0$, then there exists an orthonormal frame $(e^a_\lam)$ such that
\begin{equation}
e^0_k=0,\qq 1\le k\le n,
\end{equation}
and
\begin{equation}
e^{a'}_{k;0}=e^{a'}_{,0}-\cha k0\lam e^{a'}_\lam=0
\end{equation}
for all $1\le a'\le n$ and $1\le k\le n$.
\el
\bp
Assume that
\begin{equation}
\bar g_{00}=-w^2,
\end{equation}
then define the conformal metric
\begin{equation}
\tilde g_{\al\bet}=w^{-2}\bar g_{\al\bet}.
\end{equation}
The curves
\begin{equation}
(\ga^\al(t,x))=(t,x^i),\qq x\in\so,
\end{equation}
are then geodesics with respect to $\tilde g_{\al\bet}$. Let $(\hat e^{a'}_\lam)$, $1\le a'\le n$, be an orthonormal frame in $T^{0,1}(\so)\hra T^{0,1}(N)$ such that
\begin{equation}
\hat e^{a'}_0=0\qq\A \, 1\le a'\le n.
\end{equation}
The $\hat e^{a'}$ depend on $x=(x^i)\in \so$. Let $(\tilde e^{a'}_\lam)(t,x)$ be the solutions of the flow equations
\begin{equation}
\begin{aligned}
\frac D{dt}\tilde e^{a'}_\lam&=0,\\
\tilde e^{a'}_\lam(0,x)&=\hat e^{a'}_\lam(x),
\end{aligned}
\end{equation}
i.e., we parallel transport $\hat e^{a'}$ along the geodesics. Setting
\begin{equation}
(\tilde e^0_\lam)=(1,0,\ldots,0)
\end{equation}
the $(\tilde e^a_\lam)$ are then an orthonormal frame of $1$-forms in $(N,\tilde g_{\al\bet})$ such that the $\tilde e^a$ satisfy
\begin{equation}\lae{3.16}
\tilde e^a_{\lam:0}=0\qq\A\, 0\le a\le n,
\end{equation}
where we indicate covariant differentiation with respect to $\tilde g_{\al\bet}$ by a colon.

Define $e^a_\lam$ by
\begin{equation}
e^a_\lam=w \tilde e^a_\lam,
\end{equation}
then the $e^a_\lam$ are orthonormal frames in $(N,\bar g_{\al\bet})$. The Christoffel symbols $\cha \al\bet\ga$ \resp $\tilde \C^\ga_{\al\bet}$ are related by the formula
\begin{equation}\lae{3.18}
\begin{aligned}
\cha \al\bet\ga= \tilde \C^\ga_{\al\bet} - w^{-1} w_\al \de^\ga_\bet +w^{-1}w_\bet \de^\ga_\al -w^{-1} \check w^\ga \tilde g_{\al\bet},
\end{aligned}
\end{equation}
where 
\begin{equation}
\check w^\ga=\tilde g^{\ga\lam}w_\lam.
\end{equation}

In view of \re{3.16} we then infer
\begin{equation}
0=\tilde e^{a'}_{j:0}=\dot{\tilde e}^{a'}_j-\tilde \C^k_{0j}\tilde e^{a'}_k
\end{equation}
and we deduce further
\begin{equation}
\begin{aligned}
e^{a'}_{j;0}&=\dot w\tilde e^{a'}_j+w\dot{\tilde e}^{a'}_j-\cha 0jk w \tilde e^{a'}_k\\
&=\dot w \tilde e^{a'}_j+\tilde\C ^k_{0j}w\tilde e^{a'}_k -\cha 0jk w\tilde e^{a'}_k\\
&=0
\end{aligned}
\end{equation}
because of \re{3.18}.
\ep
Subsequently we shall always use these particular orthonormal frames.

We are now able to simplify the expressions for the spin connections
\begin{equation}
\C_\mu= -\tfrac14 E^\lam_a e^b_{\lam;\mu}\ga_a\ga^b.
\end{equation}
We have
\begin{equation}
\begin{aligned}
4\C_0&=-E^\lam_a e^b_{\lam;0}\ga_b\ga^a\\
&=-E^\lam_ae^0_{\lam;0}\ga_0\ga^a-E^\lam_ae^{b'}_{\lam;0}\ga_{b'}\ga^a\\
&=-E^0_0 e^0_{0;0}\ga_0\ga^0-E^i_{a'}e^0_{i;0}\ga_0\ga^{a'}-E^0_0e^{b'}_{0;0}\ga_{b'}\ga^0-E^i_{a'}e^{b'}_{i;0}\ga_{b'}\ga^{a'}\\
&=-E^i_{a'}e^0_{i;0}\ga_0\ga^{a'}-E^0_0e^{b'}_{0;0}\ga_{b'}\ga^0
\end{aligned}
\end{equation}
in view of \rl{3.1} and the fact that
\begin{equation}
e^0_{0;0}=0.
\end{equation}

The matrices $\ga_0\ga^{a'}$ and $\ga_{b'}\ga^0$ are hermitian, since $\ga^0$ is antihermitean, $\ga^{a'}$ hermitean and there holds
\begin{equation}
\ga_0\ga^{a'}=-\ga^{a'}\ga_0.
\end{equation}

Hence, the quadratic form
\begin{equation}
\tilde\psi E^0_a \ga^a\C_0\psi=-iE^0_0\bar\psi\C_0\psi
\end{equation}
is imaginary and will be eliminated by adding its complex conjugate. $\C_0$ can therefore be ignored which we shall indicate by writing
\begin{equation}
\C_0\simeq 0.
\end{equation}
A similar notation should apply to other terms that will be cancelled when adding the complex conjugates.

Let us consider $\C_k$:
\begin{equation}
\begin{aligned}
4\C_k&=-E^\lam_a e^b_{\lam;k}\ga_b\ga^a\\
&=-E^\lam_ae^0_{\lam;k}\ga_0\ga^a-E^\lam_ae^{b'}_{\lam;k}\ga_{b'}\ga^a\\
&=-E^0_0 e^0_{0;k}\ga_0\ga^0-E^i_{a'}e^0_{i;k}\ga_0\ga^{a'}-E^0_0e^{b'}_{0;k}\ga_{b'}\ga^0-E^i_{a'}e^{b'}_{i;k}\ga_{b'}\ga^{a'}.
\end{aligned}
\end{equation}
The first term on the right-hand side vanishes, since
\begin{equation}
e^0_{0;k}=w_k-\cha0k0w=0.
\end{equation}
Furthermore, there holds
\begin{equation}
e^0_{i;k}=-\cha ik0w=-\tfrac12\dot g_{ik}w^{-1}
\end{equation}
and
\begin{equation}
e^{b'}_{0;k}=-\cha 0kje^{b'}_j=-\tfrac12 g^{lj}\dot g_{kl}e^{b'}_j,
\end{equation}
yielding
\begin{equation}\lae{3.32}
\begin{aligned}
4\C_k&=\tfrac12 \dot g_{ik}w^{-1}E^i_{a'}\ga_0\ga^{a'}+\tfrac12 w^{-1}g^{lj}\dot g_{kl} e^{b'}_i\ga_{b'}\ga^0-E^i_{a'}e^{b'}_{i;k}\ga_{b'}\ga^{a'}\\
&=w^{-1}\dot g_{ik}E^i_{a'}\ga_0\ga^{a'}-E^i_{a'}e^{b'}_{i;k}\ga_{b'}\ga^{a'},
\end{aligned}
\end{equation}
since
\begin{equation}
\ga_0\ga^{a'}=-\ga^{a'}\ga_0.
\end{equation}
The first term on the right-hand side of \re{3.32} has to be eliminated because of the presence of $\dot g_{ik}$. To achieve this fix a Riemannian metric $\rho_{ij}=\rho_{ij}(x)\in T^{0,2}(\so)$ and define the function $\f$ by
\begin{equation}
\f=\sqrt{\frac{\det g_{ij}}{\det \rho_{ij}}}
\end{equation}
and the spinors $\chi=(\chi^i_A)$ by
\begin{equation}\lae{3.35}
\chi=\sqrt\f\psi,
\end{equation}
then
\begin{equation}
\dot \chi=\sqrt\f\dot\psi+\tfrac14 g^{ij}\dot g_{ij}\chi
\end{equation}
and
\begin{equation}
\chi_{,k}=\tfrac12 \f_k\f^{-1/2}\chi +\sqrt\f \psi_{,k}.
\end{equation}
Looking at the real part of the quadratic form
\begin{equation}
i\tilde \chi E^k_{a'}\ga^{a'}\chi_{,k}
\end{equation}
we deduce that
\begin{equation}
\chi_{,k}\simeq \sqrt\f \psi_{,k}.
\end{equation}

Moreover, we infer
\begin{equation}
\begin{aligned}
i\tilde\psi E^k_{c'}\ga^{c'}\C_k\psi&= i\bar \psi E^k_{c'}\ga^0\ga^{c'}\C_k\psi\\
&=\tfrac14 i\bar\psi E^k_{c'}E^j_{a'}w^{-1}\dot g_{jk}\ga^0\ga^{c'}\ga_0\ga^{a'}\psi\\
&\q -\tfrac14 i \bar\psi E^k_{c'}E^j_{a'}e^{b'}_{j;k}\ga^0\ga^{c'}\ga_{b'}\ga^{a'}\psi.
\end{aligned}
\end{equation}
We now observe that
\begin{equation}
\ga^0\ga^{c'}\ga_0\ga^{a'}=-\ga^0\ga_0\ga^{c'}\ga^{a'}=-\ga^{c'}\ga^{a'},
\end{equation}
hence
\begin{equation}
E^k_{c'}E^j_{a'}\ga^0\ga^{c'}\ga_0\ga^{a'}=-E^k_{c'}E^j_{a'}\ga^{c'}\ga^{a'}=-g^{jk}
\end{equation}
and we conclude
\begin{equation}
\begin{aligned}
i\tilde\psi E^\mu_c\ga^cD_\mu\psi\f&\simeq-i\bar\chi \dot\chi w^{-1}\\
&\q +i\bar\chi E^k_{c'}\ga^0\ga^{c'}\{\chi_{,k}-\tfrac14 E^j_{a'}e^{b'}_{j;k}\ga_{b'}\ga^{a'}\chi +  A_k\chi\}
\end{aligned}
\end{equation}
\br
The term in the braces is the covariant derivative of $\chi$ with respect to the spin connection $\tilde\C_k$
\begin{equation}
\tilde\C^{b'}_{ka'}=\tfrac14\tilde\om^{b'}_{ka'}=-\tfrac14 E^j_{a'}e^{b'}_{j;k}\ga_{b'}\ga^{a'}
\end{equation}
and the Yang-Mills connection $(A_\mu)$ satisfying $A_0=0$ such that
\begin{equation}\lae{3.45}
\tilde D_k\chi=\chi_{,k}+\tilde\C_k\chi+ A_k\chi.
\end{equation}
The gauge transformations for both the Yang-Mills connection as well as for the spin connection do not depend on $x^0$ but only on $x\in\so$. In case of the Yang-Mills connection this has already been proved in \cite[Lemma 2.6]{cg:uqtheory} while the proof for the spin connection $\tilde\C_k$ will be given in the next section.
\er
Summarizing the preceding results we obtain:
\bl\lal{3.3}
The Dirac Lagrangian can be expressed in the form 
\begin{equation}\lae{3.46}
\begin{aligned}
L_D&=\tfrac{i}2(\bar\chi_I\dot\chi^I-\dot{\bar\chi}^{I}\chi_{I})w^{-1}\f^{-1}+m i\bar\chi_I\ga^0\chi^I\f^{-1}\\
&\q-\tfrac{i}2\{\bar\chi_I\ga^0E^k_{a'}\ga^{a'}\tilde D_k\chi^I-\overline{\bar\chi_I\ga^0E^k_{a'}\ga^{a'}\tilde D_k\chi^I}\}\f^{-1},
\end{aligned}
\end{equation}
where $\chi$ and $\tilde D_k$ are defined in \re{3.35} \resp \re{3.45}.  
\el

\section{Quantization of the Lagrangian}\las{4}
We consider the functional
\begin{equation}\lae{4.1}
\begin{aligned}
J&=\al_M^{-1}\int_{\tilde\Om}(\bar R-2\Lam)-\int_{\tilde\Om}\tfrac14 \ga_{\bar a\bar b}\bar g^{\mu\rho_2}\bar g^{\lam\rho_1}F^{\bar a}_{\mu\rho_1}F^{\bar b}_{\rho_2\lam}\\
&\hp =\;\;\,-\int_{\tilde\Om}\{\tfrac12 \bar g^{\mu\lam}\ga_{\bar a\bar b}\F^{\bar a}_\mu\bar\F^{\bar b}_\lam+V(\F)\}\\
&\hp =\;\;+\int_{\tilde\Om}\{\tfrac12[\tilde\psi_{I}E^\mu_a\ga^a(D_\mu \psi)^{I}+\overline{\tilde\psi_{I}E^\mu_a\ga^a(D_\mu \psi)^{I}}]+m\tilde\psi_{I}\psi^{I}\},
\end{aligned}
\end{equation}
where $\al_M$ is a positive coupling constant and $\tilde\Om\Su N$.  

We use the action principle that, for an arbitrary $\tilde\Om$ as above, a solution $(A,\F,\psi,\bar g)$ should be a stationary point of the functional with respect to compact variations. This principle requires no additional surface terms for the functional.

As we proved in \cite{cg:qgravity} we may only consider metrics $\bar g_{\al\bet}$ that split with respect to some fixed globally defined time function $x^0$ such that
\begin{equation}
d\bar s^2=-w^2 (dx^0)^2+g_{ij}dx^idx^j
\end{equation}
where $g(x^0,\cdot)$ are Riemannian metrics in $\so$,
\begin{equation}
\so=\{x^0=0\}.
\end{equation}
The first functional on the right-hand side of \re{4.1} can be written in the form 
\begin{equation}\lae{5.4}
\al^{-1}_N\int_a^b\int_\Om\{\tfrac14G^{ij,kl}\dot g_{ij}\dot g_{kl}+R-2\Lam\}w\f,
\end{equation}
where
\begin{equation}
G^{ij,kl}=\tfrac12\{g^{ik}g^{jl}+g^{il}g^{jk}\}-g^{ij}g^{kl}
\end{equation}
is the DeWitt metric,
\begin{equation}
(g^{ij})=(g_{ij})^{-1},
\end{equation}
$R$ the scalar curvature of the slices
\begin{equation}
\{x^0=t\}
\end{equation}
with respect to the metric $g_{ij}(t,\cdot)$, and where we also assumed that $\tilde\Om$ is a cylinder
\begin{equation}
\tilde\Om=(a,b)\times\Om,\qq\Om\Su \so,
\end{equation}
such that $\tilde\Om\su U_k$ for some $k\in \N$, where the $U_k$ are  special coordinate patches of $N$ such that there exists a local trivialization in $U_k$ with the properties that there is a fixed  Yang-Mills connection 
\begin{equation}
\bar A=(\bar A^{\bar a}_\mu)=f_{\bar a}\bar A^{\bar a}_\mu dx^\mu
\end{equation}
satisfying
\begin{equation}
\bar A^{\bar a}_0=0\qq \text{in}\; U_k,
\end{equation}
\cf \cite[Lemma 2.5]{cg:uqtheory}. We may then  assume that the Yang-Mills connections $A=(A^{\bar a}_\mu)$ are of the form
\begin{equation}
A^{\bar a}_\mu (t,x)=\bar A^{\bar a}_\mu (0,x)+\tilde A^{\bar a}_\mu (t,x),
\end{equation}
where $(\tilde A^{\bar a}_\mu )$ is a tensor, see \cite[Section 2]{cg:uqtheory}.

The Riemannian metrics $g_{ij}(t,\cdot)$ are elements of the bundle $T^{0,2}(\so)$. Denote by $\tilde E$ the fiber bundle with base $\so$ where the fibers $F(x)$ consists of the Riemannian metrics $(g_{ij})$. We shall consider each fiber to be a Lorentzian manifold equipped with the DeWitt metric. Each fiber $F$ has dimension
\begin{equation}
\dim F=\frac{n(n+1)}2\equiv m+1.
\end{equation}
Let $(\xi^r)$, $0\le r\le m$, be  coordinates for a local trivialization such that
\begin{equation}
g_{ij}(x,\xi^r)
\end{equation}
is a local embedding. The DeWitt metric is then expressed as
\begin{equation}
G_{rs}=G^{ij,kl}g_{ij,r}g_{kl,s},
\end{equation}
where a comma indicates partial differentiation.  
In the new coordinate system the curves 
\begin{equation}
t\ra g_{ij}(t,x)
\end{equation}
can be written in the form
\begin{equation}
t\ra \xi^r(t,x)
\end{equation}
and we infer
\begin{equation}
G^{ij,kl}\dot g_{ij}\dot g_{kl}=G_{rs}\dot\xi^r\dot\xi^s. 
\end{equation}
Hence, we can express \re{5.4} as 
\begin{equation}\lae{3.49}
J=\int_a^b\int_\Om \al_n^{-1}\{\tfrac14 G_{rs}\dot\xi^r\dot\xi^sw^{-1}\f+(R-2\Lam)w\f\},
\end{equation}
where we now refrain from writing down the density $\sqrt\rho$ explicitly, since it does not depend on $(g_{ij})$ and therefore should not be part of the Legendre transformation.  Here we follow Mackey's advice in \cite[p. 94]{mackey:book} to always consider rectangular coordinates when applying canonical quantization, which can be rephrased that the Hamiltonian has to be a coordinate invariant, hence no densities are allowed.

Denoting the Lagrangian \tit{function} in \re{3.49} by $L$, we define
\begin{equation}
\pi_r= \pde L{\dot\xi^r}=\f G_{rs}\frac1{2\al_N}\dot\xi^sw^{-1}
\end{equation}
and we obtain for the Hamiltonian function $\hat H_G$
\begin{equation}\lae{5.18}
\begin{aligned}
\hat H_G&=\dot\xi^r\pde L{\dot\xi^r}-L\\
&=\f G_{rs}\big(\frac1{2\al_N}\dot\xi^rw^{-1}\big)\big(\frac1{2\al_N}\dot\xi^sw^{-1}\big) w\al_N-\al_N^{-1}(R-2\Lam)\f w\\
&=\f^{-1}G^{rs}\pi_r\pi_s w\al_N-\al^{-1}_N(R-2\Lam)\f w\\
&\equiv H_G w,
\end{aligned}
\end{equation}
where $G^{rs}$ is the inverse metric. Hence,
\begin{equation}
H_G=\al_N\f^{-1}G^{rs}\pi_r\pi_s-\al_N^{-1}(R-2\Lam)\f
\end{equation}
is the Hamiltonian that will enter the Hamilton constraint.

The Yang-Mills Lagrangian can be expressed as
\begin{equation}\lae{3.39}
L_{YM}=\tfrac12\ga_{\bar a\bar b}g^{ij}\tilde A^{\bar a}_{i,0}\tilde A^{\bar b}_{j,0}w^{-1}\f-\tfrac14 F_{ij}F^{ij}w\f.
\end{equation}
Let $E_0$ be the adjoint bundle
\begin{equation}
E_0=(S_0,\mf g,\pi,\Ad(\mc G))
\end{equation}
with base space $\so$, where the gauge transformations only depend on the spatial variables $x=(x^i)$. Then the mappings $ t\ra\tilde A^{\bar a}_i(t,\cdot)$ can be looked at as curves in $T^{1,0}(E_0)\otimes T^{0,1}(\so)$, where
the fibers of $T^{1,0}(E_0)\otimes T^{0.1}(\so)$ are the tensor products
\begin{equation}
\mf g\otimes T^{0,1}_x(\so),\qq x\in \so,
\end{equation}
which are vector spaces equipped with metric
\begin{equation}
\ga_{\bar a\bar b}\otimes g^{ij}. 
\end{equation}
For our purposes it is more convenient to consider the fibers to be Riemannian manifolds endowed with the above metric. Let $(\zeta^p)$, $1\le p\le n_1n$, where $n_0=\dim\mf g$, be  local coordinates and
\begin{equation}
(\zeta^p)\ra \tilde A^{\bar a}_i(\zeta^p)\equiv \tilde A(\zeta)
\end{equation}
be a local embedding, then the metric has the coefficients
\begin{equation}
G_{pq}=\spd{\tilde A_p}{\tilde A_q}=\ga_{\bar a\bar b}g^{ij}\tilde A^{\bar a}_{i,p}\tilde A^{\bar b}_{j,q}.
\end{equation}
Hence, the Lagrangian $L_{YM}$ in \re{3.39} can be expressed in the form
\begin{equation}
L_{YM}=\tfrac12G_{pq}\dot\zeta^p\dot\zeta^qw^{-1}\f-\tfrac14F_{ij}F^{ij}w\f
\end{equation}
and we deduce
\begin{equation}
\tilde\pi_p=\pde{L_{YM}}{\dot\zeta^p}=G_{pq}\dot\zeta^qw^{-1}\f
\end{equation}
yielding the Hamilton function
\begin{equation}
\begin{aligned}
\hat H_{YM}&=\pi_p\dot\zeta^p-L_{YM}\\
&=\tfrac12 G_{pq}(\dot\zeta^pw^{-1}\f)(\dot\zeta^qw^{-1}\f)w\f^{-1}+\tfrac14F_{ij}F^{ij}w\f\\
&=\tfrac 12G^{pq}\tilde\pi_p\tilde\pi_qw\f^{-1}+\tfrac14F_{ij}F^{ij}w\f\\
&\equiv H_{YM}w.
\end{aligned}
\end{equation}
Thus, the Hamiltonian that will enter the Hamilton constraint equation is
\begin{equation}\lae{3.51}
H_{YM}=\tfrac 12\f^{-1}G^{pq}\tilde\pi_p\tilde\pi_q+\tfrac14F_{ij}F^{ij}\f.
\end{equation}

Using the Hamilton gauge as before
the Higgs Lagrangian can be written as
\begin{equation}
L_H=\tfrac12 \ga_{\bar a\bar b}\F^{\bar a}_{,0}\F^{\bar b}_{,0}w^{-1}\f-\tfrac12g^{ij}\ga_{\bar a\bar b}\F^{\bar a}_i\F^{\bar b}_jw\f-V(\F)w\f
\end{equation}
which we have to use for the Legendre transformation; here, we also assume without loss of generality that the coefficients of $\F$ are real without changing the notation or the indices though the number of components have doubled. However, later, when we define the fiber bundle, we shall use the correct real dimensions, i.e., we shall use $\mf g\times \mf g$ instead of $\mf g$. Before applying the Legendre transformation we again consider the vector space $\mf g$ to be a Riemannian manifold with metric $\ga_{\bar a\bar b}$. The representation of $\F$ in the form $(\F^{\bar a})$ can be looked at to be the representation in a local coordinate system $(\Theta^{\bar a})$.

Let us define
\begin{equation}
p_{\bar a}=\pde{L_H}{\dot\F^{\bar a}},\qq\dot\F^{\bar a}=\F^{\bar a}_{,0},
\end{equation}
then we obtain the Hamiltonian
\begin{equation}
\begin{aligned}
\hat H_H&=p_{\bar a}\dot\F^{\bar a}-L_H\\
&=\tfrac12\ga_{\bar a\bar b}(\dot\F^{\bar a}w^{-1}\f)(\dot\F^{\bar b}w^{-1}\f)w\f^{-1}+\tfrac12 g^{ij}\ga_{\bar a\bar b}\F^{\bar a}_i\F^{\bar b}_jw\f+V(\F)w\f\\
&\equiv H_Hw.
\end{aligned}
\end{equation}
Thus, the Hamiltonian which will enter the Hamilton constraint is
\begin{equation}\lae{4.10}
H_H=\tfrac12\f^{-1}\ga^{\bar a\bar b}p_{\bar a}p_{\bar b}+\tfrac12 g^{ij}\ga_{\bar a\bar b}\F^{\bar a}_i\F^{\bar b}_j\f+V(\F)\f.
\end{equation}

The spinorial variables  $\chi^I_A$ are anticommuting Grassmann variables. They  are elements of a  Grassmann algebra with involution, where the involution corresponds to the complex conjugation and will be denoted by a bar.

The $\chi^{I}_A$ are complex variables and we define its real \resp imaginary parts as
\begin{equation}
\xi^{I}_A=\tfrac1{\sqrt 2}(\chi^{I}_A+\bar\chi^{I}_A)
\end{equation}
\resp
\begin{equation}
\h^{I}_A=\tfrac1{\sqrt 2 i}(\chi^{I}_A-\bar\chi^{I}_A).
\end{equation}
Then,
\begin{equation}\lae{4.6.1}
\chi^{I}_A=\tfrac1{\sqrt 2}(\xi^{I}_A+i\h^{I}_A)
\end{equation}
and
\begin{equation}\lae{4.7.1}
\bar\chi^{I}_A=\tfrac1{\sqrt 2}(\xi^{I}_A-i\h^{I}_A).
\end{equation}

With these definitions we obtain
\begin{equation}\lae{4.9}
\frac{i}2(\bar\chi_{I}\dot\chi^{I}-\bar{\dot\chi}^{I}\chi_{I})=\frac{i}2(\xi^A_{I}\dot\xi^{I}_A+\h^A_{I}\dot\h^{I}_A).
\end{equation}

Casalbuoni  quantized a Bose-Fermi system in \cite[section 4]{casalbuoni:fermi} the results of which can be applied to spin $\frac12$ fermions. The Lagrangian in \cite{casalbuoni:fermi} is the same as the main part  our Lagrangian in \fre{3.46}, and the left derivative is used in that paper, hence we are using left derivatives as well such that  the conjugate momenta of the odd variables are, e.g.,
\begin{equation}
\pi^A_{I}=\frac{\pa L}{\pa \dot\xi^{I}_A}=-\frac i2\xi^A_{I},
\end{equation}
and thus the conclusions in \cite{casalbuoni:fermi} can be applied. 

The Lagrangian has been expressed in real variables---at least the important part of it---and it follows that the odd variables $\xi^{I}_A,\h^{I}_A$ satisfy, after introducing anticommutative Dirac brackets as in \cite[equ. (4.11)]{casalbuoni:fermi}, 
\begin{equation}
\{\xi_{I}^A,\xi^{J}_B\}^*_+=-i\de^{J}_I\de_{B}^A,
\end{equation}
\begin{equation}
\{\h_{I}^A,\h^{J}_B\}^*_+=-i\de^{J}_I\de_{B}^A,
\end{equation}
and
\begin{equation}
\{\xi_{I}^A,\h^{J}_B\}^*_+=0,
\end{equation}
\cf \cite[equ. (4.19)]{casalbuoni:fermi}.

In view of \re{4.6.1}, \re{4.7.1} we then derive
\begin{equation}
\{\bar\chi_{I}^A,\chi^{J}_B\}^*_+=-i\de^{J}_I\de_{B}^A.
\end{equation}

Canonical quantization---with $\bar h=1$---then requires that the corresponding operators $\hat\chi^{I}_A, \hat{\bar\chi}_{J}^B$ satisfy the anticommutative rules 
\begin{equation}\lae{4.15.1}
[\hat\chi^{I}_A,\hat{\bar\chi}_{J}^B]_+=i\{\chi^{I}_A,\bar\chi_{J}^B\}^*_+=\de^{I}_J\de_{A}^B
\end{equation}
and
\begin{equation}
[\hat{\bar\chi}_{I}^A,\hat{\bar\chi}_{J}^B]_+=[\hat\chi^{I}_A,\hat{\chi}^{J}_B]_+=0,
\end{equation}
\cf \cite[equ. (3.10)]{casalbuoni:odd} and \cite[equ. (5.17)]{casalbuoni:fermi}.

We shall realize these quantum rules in a suitable Grassmann algebra which will be equipped with a natural scalar product.

Let $g_{ij}$ be an arbitrary but fixed Riemannian metric in $\so$ and define $N=I\times \so$ to be the Lorentz manifold endowed with the product metric
\begin{equation}
d\bar s^2=-(dx^0)^2+g_{ij}(x) dx^idx^j.
\end{equation}
Let $e^{a'}_i$  be an orthonormal frame on $\so$ and $E^i_{a'}$  its inverse. This orthonormal frame can be lifted to $N$ by setting
\begin{equation}
e^a_0=\de^a_0\qq\wed\qq e^0_a=\de^0_a.
\end{equation}
The orthonormal frame $e^a_\mu$ then satisfies the conditions in \frl{3.1}.

Let $\C_\mu$ be the corresponding spin connection, then
\begin{equation}\lae{4.49}
\C_0=0,
\end{equation}
and hence, the spinorial gauge transformations only depend on $x\in\so$ and not on $x^0$, since any vielbein is supposed to satisfy \re{4.49}. 

Now, let $\chi^I_A$ be a spinor field in $N$, define
\begin{equation}
\psi^I_A(x)=\chi^I_A(0,x)
\end{equation}
and let $\psi^I_A(t,x)$ be the solution of the flow
\begin{equation}
\begin{aligned}
\frac D{dt}\psi^I_A&\equiv E^\mu_0D_\mu\psi^I_A=0,\\
\psi^I_A(0,x)&=\psi^I_A(x).
\end{aligned}
\end{equation}
Then, $\psi^I_A$ is a spinor field in $N$ satisfying
\begin{equation}
\psi^I_A(x^0,x)=\psi^I_A(x),
\end{equation}
because
\begin{equation}
\begin{aligned}
0=\frac D{dt}\psi&=\dot\psi+\C_0\psi+A_0\psi\\
&=\dot\psi.
\end{aligned}
\end{equation}

In the following we shall only consider spinor fields with this property   calling them spinor fields defined in $\so$ and we shall mostly use the symbol $\chi^I_A=\chi^I_A(x)$. We treat the components $\chi^I_A$ as Grassmann variables and define:
\bd
Let $\chi^i_A$ be Grassmann variables, then we define
\begin{equation}
\mc P=\mc P(\chi^I_A)
\end{equation}
to be the space of polynomial functions
\begin{equation}\lae{4.55}
u=a_0+\sum_{k,m}a^{A_1\cdots A_m}_{I_1\cdots I_k}\chi^{I_1\cdots I_k}_{A_1\cdots A_m},
\end{equation}
where the coefficients are complex numbers being asymmetric in their indices. The indices are also supposed to satisfy
\begin{equation}\lae{4.56}
A_1<\cdots<A_m\q\wed\q I_1<\cdots<I_k.
\end{equation}
If the usual summation convention is supposed to be implemented, i.e., if the stipulation \re{4.56}  is dropped, then
\begin{equation}
u=a_0+\sum_{k,m}\frac1{k!m!}a^{A_1\cdots A_m}_{I_1\cdots I_k}\chi^{I_1\cdots I_k}_{A_1\cdots A_m}.
\end{equation}
However, we prefer to use the representation \re{4.55} with the implicit understanding of \re{4.56}.
\ed
\br
The vector space $\mc P(\chi^I_A)$ is a Grassmann algebra with base vectors
\begin{equation}\lae{4.58}
\{\mathds{1}, \chi^{I_1\cdots I_k}_{A_1\cdots A_m}:A_1<\cdots<A_m\q\wed\q I_1<\cdots<I_k\},
\end{equation}
where $\mathds1$ is the symbol for the unit element. If we define
\begin{equation}\lae{4.59}
\bar\chi^A_I=\frac{\pa}{\pa\chi^I_A}
\end{equation}
to be the left derivative and
\begin{equation}
\bar u=\bar a_0+\sum_{k,m}\bar a_{A_1\cdots A_m}^{I_1\cdots I_k}\bar\chi_{I_1\cdots I_k}^{A_1\cdots A_m},
\end{equation}
where the indices of the coefficients are raised or lowered with the help of the metrics $\de_{AB}$ \resp $\de_{IK}$, then $\bar u$ is a linear operator in $\mc P$. 

Stipulating that the base vectors in \re{4.58} are orthonormal we can define a Hermitian scalar product in $\mc P$ which is antihermitian in the first argument.

Any function $u\in\mc P$ can also be viewed as a linear operator by defining
\begin{equation}\lae{4.62}
uv\equiv u(v)=uv,
\end{equation}
where the right-hand side is the Grassmann product.
\er
\bl\lal{4.3}
The operators $\chi^I_A$ and $\bar\chi^A_I$ satisfy the anticommutation rules
\begin{equation}
[\bar\chi^A_I,\chi^K_B]_+=\de^A_B\de^K_I
\end{equation}
and in addition
\begin{equation}\lae{4.64}
\bar\chi^A_I=(\chi^I_A)^*,
\end{equation}
where the star indicates the adjoint operator.
\el
The proof is elementary. 

Spinorial or Yang-Mills gauge transformations then induce unitary transformations in $\mc P$.

The following lemma is worth noting:
\bl
Let $\spd\cdot\cdot$ be a Hermitian scalar product in $\mc P$ which is antihermitian in the first argument and satisfies \re{4.64}, or equivalently,
\begin{equation}
\big(\frac{\pa}{\pa\chi^I_A}\big)^*=\chi^I_A
\end{equation}
as well as
\begin{equation}
\spd{\mathds {1}}{\mathds {1}}=1,
\end{equation}
then the base vectors in \re{4.58} are orthonormal, hence it is uniquely determined.
\el
The proof is an easy exercise.

After having realized the quantization rules \fre{4.15.1} in the Grassmann algebra $\mc P$, let us look at the spinorial Hamilton function and its corresponding Hamilton operator after quantization.

From \frl{3.3} we deduce that the spinorial Hamilton function is equal to
\begin{equation}\lae{4.65}
\begin{aligned}
\hat H_D&=
\tfrac{i}2\{\bar\chi_I\ga^0E^k_{a'}\ga^{a'}\tilde D_k\chi^I-\overline{\bar\chi_I\ga^0E^k_{a'}\ga^{a'}\tilde D_k\chi^I}\}w\\
&\q-m i\bar\chi_I\ga^0\chi^Iw\\
&\equiv H_Dw.
\end{aligned}
\end{equation}
$H_D$ is the Hamilton function which has to be quantized. By applying the definitions in \re{4.59}, \re{4.62} and the results in \rl{4.3} it is obvious that $H_D$ can looked at as a self-adjoint operator in the finite dimensional Hilbert space $\mc P$ without changing its notation.

Defining $H_D$ to be an element of $L(\mc P,\mc P)$ is fairly straight-forward---only the transformation of the covariant derivative, or more precisely,  of the partial derivative
\begin{equation}
\frac{\pa}{\pa x^k}\chi^I_A
\end{equation}
requires some consideration.

For simplicity let us drop the spinor index $A$ such that we only consider the Grassmann variables $\chi^I$. To express their partial derivatives let $\psi^K$ be Grassmann variables which do not depend on $x$, then
\begin{equation}
\chi^I=a^I_k\psi^K,
\end{equation}
where $a^I_K=a^I_K(x)$. Let
\begin{equation}
(\hat a^K_I)=(a^I_K)^{-1},
\end{equation}
then
\begin{equation} 
\begin{aligned}
\chi^I_{,k}&=a^I_{K,k}\psi^K\\
&=a^I_{K,k}\hat a^K_M\chi^M
\end{aligned}
\end{equation}
and we immediately infer how $\chi^I_{,k}$ can be transformed to be a linear operator in $\mc P$.

Combining the four Hamilton functions in \re{5.18}, \re{3.51}, \re{4.10} and \re{4.65} the Hamilton constraint has the form
\begin{equation}
\begin{aligned}
H&=H_G+H_{YM}+H_H+H_D=0\\
&\equiv\tilde H+H_D,
\end{aligned}
\end{equation}
where
\begin{equation}
H=H(\xi^r,\zeta^p,\Theta^{\bar a},\tilde\Theta^{\bar b},\pi_r,\tilde\pi_q,p_{\bar c}, \tilde p_{\bar d},\chi^I_A,\bar\chi_J^B).
\end{equation}
Here, $(\xi^r,\zeta^p,\Theta^{\bar a},\tilde\Theta^{\bar b})$ are local sections of a bundle $E$ with base space $\so$ and fibers
\begin{equation}\lae{4.72}
\mc F(x)=F(x)\times(\mf g\otimes T^{0,1}_x(\so))\times\mf g\times\mf g.
\end{equation}
Applying canonical quantization by setting $\hbar =1$ we replace
\begin{equation}
\pi_r=\pi_r(x)\ra \frac1i\frac{\pa}{\pa\xi^r(x)}
\end{equation}
and similarly for the other conjugate momenta $\tilde \pi_q$, $p_{\bar c}$, and $\tilde p_{\bar d}$, while the conjugate momentum $\bar\chi^A_I$ is being replaced by the left derivative with respect to the Grassmann variable $\chi^I_A$ as described previously.

The Hamiltonian $\tilde H$ will be transformed to a normally hyperbolic differential operator in the bundle $E$ acting only in the fibers where the fibers in \re{4.72} are equipped with the Lorentzian metric
\begin{equation}\lae{4.74}
G=\f\diag(\al_N^{-1}G_{rs}, 2G_{pq},2\ga_{\bar a\bar b},2\ga_{\bar c\bar d}).
\end{equation}
The fibers are then globally hyperbolic spacetimes as we proved in \cite[Theorem 4.1]{cg:uqtheory}.

Let $\hat E$ be the vector bundle with base space $E$ and fiber
\begin{equation}\lae{4.77}
\hat{\mc  F}=T^{0,2}_{x,\tup{ symm}}(\so)\times(\mf g\otimes T^{0,1}_x(\so))\times\mf g\times \mc P,
\end{equation}
where $\mc P$ is the complex Hilbert space discussed above. Writing the elements $u$ of the fiber in coordinates we obtain
\begin{equation}
u=(f_{ij},A^{\bar a}_m,\F^{\bar b},w),
\end{equation}
where $w\in\mc P$,  the components $\F^b$ are complex, while $f_{ij}$ and $A^a_m$ are real. We consider $\hat{\mc  F}$ to be a real vector space and define the real scalar product
\begin{equation}\lae{4.79}
\begin{aligned}
\spd u{\tilde u}&=(\rho^{ik}\rho^{jl}+\rho^{il}\rho^{jk})f_{ij}\tilde f_{kl}+\ga_{\bar a\bar b}\rho^{pq}A^{\bar a}_p\tilde A^{\bar b}_q\\
&\qq+\rec (\ga_{\bar a\bar b}\F^{\bar a}\bar{\tilde\F}^{\bar b})+\rec \spd w{\tilde w},
\end{aligned}
\end{equation}
where $\rho_{ij}$ is the fixed metric in $\so$ which is used to define $\f$. $\rho_{ij}$ should be chosen such that it has an interesting group of isometries, since these isometries and the gauge transformations act naturally on the elements of the fiber such that they are isometries with respect to the scalar product above. 

After quantization the Hamilton function $H=\tilde H+H_D$ is replaced by a normally hyperbolic differential operator, also denoted by $H$, which can be looked at as a map from the sections $C^\un_c(E,\hat E)$ into itself. The linear map $H_D$ is part of the zero order term of $H$. $H_D$ can be trivially extended to act in the fibers.

The Wheeler-DeWitt equation has the form
\begin{equation}
Hu=0
\end{equation}
with $u\in C^\un(E,\hat E)$.
\bl\lal{4.5}
The isometries of $\rho_{ij}$ and the Yang-Mills and spin  gauge transformations commute with $H$, hence the kernel of $H$ is invariant under these actions.
\el

\bp
It suffices to prove the claim for spin gauge transformations. Let $\F$ be such a transformation, then $\F$ defines a new variable $\psi_A$ in the Gra{\ss}mann algebra $\mc P$---note that we only consider one index to simplify the notation. When we look at $\chi_A$ to be a  vector, then
\begin{equation}
\psi_A=\F(\chi_A)
\end{equation}
and $\F$ is unitary for these particular basis vectors. We extend $\F$ to the other basis vectors by setting
\begin{equation}
\F(\mathds 1)=\F(\mathds 1)
\end{equation}
and
\begin{equation}
\F(\chi_A\chi_B)=\F(\chi_A)\F(\chi_B)
\end{equation}
and similarly in case of more factors. Hence, we obtain
\begin{equation}
\F(\chi_A\chi_B)=\psi_A\psi_B.
\end{equation}
$\F$ is then unitary in $\mc P$ and we obtain an orthonormal basis by simply replacing the $\chi$'s by the $\psi$'s.

Now, to prove that $\F$ commutes with the Hamiltonian $H$ it suffices to only consider the Dirac Hamiltonian $H_D$ and to prove that $\F$ commutes with $H_D$. $H_D$ is an invariant with respect to gauge transformations, i.e.
\begin{equation}
\F(H_D)=H_D.
\end{equation}
Let us prove the commutation claim only in case of the base vectors
\begin{equation}
\chi_A\chi_B,
\end{equation}
then
\begin{equation}
\begin{aligned}
H_D\F(\chi_A\chi_B)&=H_D\psi_A\psi_B\\
&=\F(H_D)\psi_A\psi_B\\
&=\F(H_D\chi_A\chi_B)
\end{aligned}
\end{equation}
because of the symmetry with respect to the variables $\chi_A$ and $\psi_A$, proving the lemma.
\ep

\br
Since $\mc P$ is a Gra{\ss}mann algebra with conjugation we also have
\begin{equation}
\overline{\F(\chi_A)}=\F(\bar\chi_A)
\end{equation}
as one easily checks by using the fact that
\begin{equation}
\bar\chi_A=\chi_A^*,
\end{equation}
\cf \re{4.64}.
\er
Let each fiber $\mc F(x)$ of $E$ be equipped with the Lorentz metric $G_{ab}$ in \re{4.74}, then there exists a natural measure on $E$ and we can define a scalar product in $C^\un_c(E,\hat E)$ by setting
\begin{equation}
\spd uv_{\hat E}=\int_\so\int_{\mc F(x)}\spd uv,
\end{equation}
where $\spd uv$ is the scalar product in \re{4.79}. With respect to this scalar product $H$ is symmetric, i.e.,
\begin{equation}
\spd{Hu}v_{\hat E}=\spd u{Hv}_{\hat E}\qq\A\, u,v\in C^\un_c(E,\hat E),
\end{equation}
since $Hu$ can be expressed as
\begin{equation}
Hu=-\D u+cu.
\end{equation}
The self-adjoint operator $H_D$ is part of the coefficient $c$. Let us emphazise that, apart from $H_D$, $H$ is acting diagonally on each component of $u$.

\section{The method of second quantization}\las{5}

In the previous sections we used canonical quantization to quantize a classical system leading to the Wheeler-DeWitt equation which can be solved subject to Cauchy conditions. Indeed the solution space  will be infinite dimensional.

To describe the existence results and the necessary techniques we first need a definition:
\bd
A Cauchy hypersurface in the bundle $E$ is a subbundle $M$ with same base space $\so$ such that each fiber $M(x)$ is a Cauchy hypersurface in the corresponding fiber $\mc F(x)$ of $E$. 
\ed
The non-homogeneous Cauchy problems for the Hamilton operator $H$ are then uniquely solvable:
\bt\lat{5.2}
Let $M\su  E$ be a Cauchy hypersurface with future directed normal $\nu$, $H$ the Hamilton operator, $u_0$, $u_1$ \resp $f$ sections in $C^\un_c(M,\hat E)$ \resp $C^\un_c(E,\hat E)$, then the Cauchy problem
\begin{equation}
\begin{aligned}
Hu&=f,\\
\fv uM&=u_0,\\
\fv{u_\al\nu^\al}M&=u_1,
\end{aligned}
\end{equation}
has a unique solution $u\in C^\un(E,\hat E)$ such that
\begin{equation}\lae{5.2}
\supp u\su J^E(K)=\uuu_{x\in \so} J^{F(x)}(K(x)),
\end{equation}
where
\begin{equation}
K=\supp u_0\,\uu\,\supp u_1\,\uu\,\supp f
\end{equation}
and
\begin{equation}
K(x)=K\ii \pi^{-1}(x),\qq x\in \so,
\end{equation}
 $\pi$ is the projection from $E$ to $\so$. Furthermore,
\begin{equation}
J^E(K)=J^E_+(K)\,\uu\,J^E_-(K)
\end{equation}
and
\begin{equation}
J^E_\pm(K)=\uuu_{x\in \so} J^{F(x)}_\pm(K(x));
\end{equation}
these are the points that can be reached by causal curves starting in $K$. Moreover, $u$ depends continuously on the data $(u_0,u_1,f)$ with corresponding estimates, namely, for any compact sets $K, K_1\su N$ and $K_0\su M$ and any $m\in\N$ there exists $m'\in\N$ and a constant $c=c(m, m', K,K_0,K_1)$ such that 
\begin{equation} 
\abs u_{m,K}\le c\{\abs{u_0}_{m',K_0}+\abs {u_1}_{m',K_0}+\abs f_{m',K_1}\},
\end{equation}
where $u$ is a solution of the Cauchy problem and $u_0$, $u_1$ and $f$ have support in the respective sets $K_0$ and $K_1$. 
\et

A proof is given in \cite[Theorem 5.4]{cg:qgravity} based on the results in \cite[Theorem 3.2.11, Theorem 3.2.12]{baer:book}. Our former proof only considered functions in $C^\un(E,\Cc)$ but it is also valid in the more general setting when $u\in C^\un(E,\hat E)$.

\br\lar{5.3}
The solutions $u$ in the preceding theorem do not have compact support in $E$, but from \re{5.2} we deduce that their support is spacelike compact,  since the fibers are globally hyperbolic. We use the notation $C^\un_{sc}(E,\hat E)$ for the set of all such $u\in C^\un(E,\hat E)$ for which there exists a compact subset $K\su E$ such that
\begin{equation}
\supp u\su J^E(K),
\end{equation}
\cf the corresponding definition in \cite[Definition 3.4.5]{baer:book}. Sections with spacelike compact support have the important property that the intersection of $\supp u$ with any Cauchy hypersurface is compact, \cf \cite[Corollary A.5.4]{baer:book}.
\er

The bosonic and fermionic components in the bundle $\hat E$ have been treated equally so far. However, in order to achieve a second quantization we have to define a CCR representation for the bosonic  and a CAR representation for the fermionic part. Therefore, let us split the fiber $\hat{\mc F}$ into a direct sum
\begin{equation}
\hat{\mc F}=\hat{\mc F}_{\tup{bose}}\oplus \hat{\mc F}_{\tup{ferm}}
\end{equation}
which is also an orthogonal sum for the scalar product in \fre{4.79}. The bundle $\hat E$ splits accordingly
\begin{equation}
\hat E=\hat E_{\tup{bose}}\oplus \hat E_{\tup{ferm}}\equiv \hat E_1\oplus \hat E_2,
\end{equation}
where all bundles have the common base space $E$. The bundles $\hat E_i$ are invariant under the Hamilton operator $H$, i.e., a solution $u$ of the Wheeler-DeWitt equation can be written in the form
\begin{equation}
u=u_1\oplus u_2,
\end{equation}
where each $u_i\in C^\un_{\tup{sd}}(E,\hat E_i)$ satisfies the equation
\begin{equation}
Hu_i=0.
\end{equation}

From \rt{5.2} we deduce that there exist the advanced and retarded Green distributions $G_+$ and $G_-$ for $H$ such that
\begin{equation}
G_\pm:C^\un_c(E,\hat E)\ra C^\un(E,\hat E)
\end{equation}
\begin{equation}\lae{5.10}
H\circ G_\pm=G_\pm\circ \fv H{C^\un_c(E,\hat E)}=\id_{C^\un_c(E,\hat E)}
\end{equation}
\begin{equation}\lae{5.11}
\supp \,(G_+ u)\su J^E_+(\supp u)\qq\A\, u\in C^\un_c(E,\hat E)
\end{equation}
and
\begin{equation}\lae{5.12}
\supp\,(G_- u)\su J^E_-(\supp u)\qq\A\, u\in C^\un_c(E,\hat E).
\end{equation}
We note that the Dirac Hamiltonian $H_D$ only acts in the fermionic case  non-trivially. In $\hat E_1$ there holds
\begin{equation}
H_D=0
\end{equation}
by definition.

We shall first construct a CCR representation or a Weyl system for $\hat E_1$. For simplicity we refer to sections in $C^\un(E,\hat E_1)$ by using the symbols $u,v$, etc.\ dropping the index $1$.

There are two ways to construct a Weyl system given a formally self-adjoint normally hyperbolic operator in a globally hyperbolic spacetime  or, in our case, in the bundle $E$. One possibility is   to define a symplectic vector space
\begin{equation}\lae{5.13}
V=C^\un_c(E,\hat E_1)/N(G),
\end{equation}
where $G$ is the Green's distribution 
\begin{equation}
G=G_+-G_-.
\end{equation}
Since
\begin{equation}\lae{5.15}
G^*=-G
\end{equation}
the bilinear form 
\begin{equation}\lae{5.16}
\om=\int_\so\int_{\mc F}\spd{u}{Gv}\qq u,v\in V
\end{equation}
is skew-symmetric,  non-degenerate by definition and hence symplectic, and then there is a canonical way to construct a corresponding Weyl system.

The second method is to use a Cauchy hypersurface to define a quantum field in Fock space. Let us start with this method keeping in mind that our bundle is a real vector bundle.

First we need the following lemma which was proved in \cite[Lemma 6.1]{cg:qgravity} when $u,v$ are test functions but the proof is also valid in the more general case when $u,v$ are sections with values in a vector space with a scalar product. 
\bl
Let $M$ be a Cauchy hypersurface in $E$, then 
\begin{equation}\lae{5.17}
\begin{aligned}
\int_\so\int_{\mc F}\spd u{Gv}=\int_\so\int_M\{\spd{D_\nu(Gu)}{Gv}-\spd{Gu}{D_\nu(Gv)}\}
\end{aligned}
\end{equation}
for all $u,v\in C^\un_c( E,\hat E_1)$,  where $\nu$ is the future normal to $M$ and the scalar product is the standard scalar product in $\hat E_1$.
\el 
We now define the  complex Hilbert space $H_{M}$ which is used to construct the symmetric Fock space, namely, we set
\begin{equation}
H_{M}=L^2(M,\hat E_1)\otimes\Cc
\end{equation}
to be the complexification of the real vector space $L^2(M,\hat E_1)$ with the complexified scalar product
\begin{equation}
\spd uv_{M}=\int_\so\int_M\spd uv_\Cc,
\end{equation}
where $\spd\cdot\cdot_\Cc$ is the complexification of $\spd\cdot\cdot$ in $\hat{\mc  F}_1\otimes\Cc$.

We denote the symmetric Fock space by $\mc F(H_{M})$. Let $\Theta$ be the corresponding Segal field. Since $G^*=-G$ we deduce from  \re{5.11},  \re{5.12} and \rr{5.3} 
\begin{equation}
\fv {G^*u}{M}\in C^\un_c(M, \hat E_1)\su H_{M}.
\end{equation}
We can therefore define
\begin{equation}\lae{5.21}
\F_{M}(u)=\Theta(i\fv {(G^*u)}{M}-D_\nu\fv{(G^*u)}{M}).
\end{equation}
From the proof of \cite[Lemma 4.6.8]{baer:book} we conclude that the right-hand side of  \re{5.21} is an essentially self-adjoint operator in $\mc F(H_{M})$. We therefore call the map $\F_{M}$ from $C^\un_c(E, \hat E_1)$ to the set of self-adjoint operators in $\mc F(H_{M})$ a quantum field for $H$ defined by $M$. 

\bl
The quantum field $\F_{M}$ satisfies the equation
\begin{equation}
H\F_{M}=0
\end{equation}
in the distributional sense, i.e.,
\begin{equation}
\spd{H\F_{M}}u=\spd{\F_{M}}{Hu}=\F_{M}(Hu)=0\q\A\, u\in C^\un_c(E, \hat E_1).
\end{equation}
\el
\bp
In view of \re{5.10} there holds
\begin{equation}
G^*Hu=0\qq\A\, u\in C^\un_c(E,\hat E_1).
\end{equation}
\ep

With the help of the quantum field $\F_M$ we shall construct a Weyl system and hence a CCR representation of the symplectic vector space $(V,\om)$ which we defined in \re{5.13} and \re{5.16}.

From \re{5.21} we conclude  the commutator relation
\begin{equation}
[\F_M(u),\F_M(v)]=i\imc\spd{iG^*u-D_\nu(G^*u)}{iG^*v-D_\nu(G^*v)}_MI,
\end{equation}
for all $u,v\in C^\un_c(E,\hat E_1)$, \cf \cite[Proposition 5.2.3]{robinson:book2}, where both sides are defined in the algebraic Fock space $\mc F_{\tup{alg}}(H_M)$. 

On the other hand
\begin{equation}\lae{5.26}
\begin{aligned}
\imc\spd{iG^*u-D_\nu(G^*u)}{iG^*v-D_\nu(G^*v)}_M&\\
&\msp[-350]=-\imc\spd{iG^*u}{D_\nu(G^*v)}_M-\imc\spd{D\nu(G^*u)}{iG^*v}_M\\
&\msp[-350]=\int_\so\int_M\{\spd{G^*u}{D_\nu(G^*v)}-\spd{D_\nu(G^*u)}{G^*v}\}\\
&\msp[-350]=\int_\so\int_{\mc F}\spd u{Gv}
\end{aligned}
\end{equation}
in view of \re{5.15} and \re{5.17}.

As a corollary we conclude
\begin{equation}\lae{5.27}
[\F_M(u),\F_M(v)]=i\int_\so\int_M\spd u{Gv}I\q\A\, u,v\in C^\un_c(E,\hat E_1).
\end{equation}

From \cite[Proposition 5.2.3]{robinson:book2} and \re{5.26} we immediately infer
\bt\lat{5.6}
Let $(V,\om)$ be the symplectic vector space in \re{5.13} and \re{5.16} and denote by $[u]$ the equivalence classes in $V$, then
\begin{equation}
W([u])=e^{i\F_M(u)}
\end{equation}
defines a Weyl system for $(V,\om)$, where $\F_M(u)$ is now supposed to be the closure of $\F_M(u)$ in $\mc F(H_M)$, i.e., $\F_M(u)$ is a self-adjoint operator. The Weyl system generates a $C^*$-algebra with unit which we call a CCR representation of $(V,\om)$.
\et
\br\lar{5.7}
Since all CCR representations of $(V,\om)$ are $^*$-isomorphic, where the isomorphism maps Weyl systems to Weyl systems, \cf \cite[Theorem 5.2.8]{robinson:book2},  this especially applies  to the CCR representations corresponding to different Cauchy hypersurfaces $M$ and $M'$, i.e., there exists a $^*$-isomorphism $T$ such that
\begin{equation}
T(e^{i\F_M(u)})=e^{i\F_{M'}(u)}\qq \A\, [u]\in V.
\end{equation}
\er
\bl 
The transformations $A$ in \frl{4.5} are also symplectic transformations for the symplectic form $\om$ defined in \re{5.16}, i.e.,
\begin{equation}
\om(u,v)=\om(Au,Av)\qq\A\, u,v\in V.
\end{equation}
\el
\bp
Let $F_+$, $F_-$ be the fundamental solutions of the hyperbolic operator $H$, where we suppress the dependence on a base point. Since $A$ commutes with $H$ it follows immediately from the definition of $F_{\pm}$ that $A$ also commutes with $F_\pm$ and hence with $G_+$, $G_-$ and with $G$. The result is then due to the fact that $A$ is also an isometry for the scalar product in \fre{4.79}.
\ep

Let us conclude this section with the following important theorem:
\bt
Let $H$ and $G$ be as above and define 
\begin{equation}
N(H)=\set{u\in C^\un_{\tup{sc}}(E, \hat E)}{Hu=0},
\end{equation}
then
\begin{equation}
N(H)=R(G)
\end{equation}
and
\begin{equation}
N(G)=R(H),
\end{equation}
where $R(G)$ \resp $R(H)$ are the images of $C^\un_c(E,\hat E)$ under the respective maps.
\et
The proof of this theorem is an adaption of the proof of the corresponding result in \cite[Theorem 3.4.7]{baer:book} when $E$ is not a bundle but a globally hyperbolic manifold.

\section{The CAR representation}\las{6}
Let us now consider the fermionic bundle $\hat E_2$. Its fibers are the Gra{\ss}mann algebra $\mc P$ which is also a complex Hilbert space with hermitian form
\begin{equation}
\spd uv=\spd uv_{\mc P}
\end{equation}

We fix a Cauchy hypersurface $M\su E$, which is a subbundle and define the complex Hilbert space
\begin{equation} 
\tilde {\mc H}=\tilde{\mc H}_M=L^2(M,\hat M\times \hat M),
\end{equation}
where $\hat M$ is the corresponding subbundle in $\hat E_2$. For elements $u=(u_1,u_2)$ and $v=(v_1,v_2)$ in $\tilde{\mc H}_M$ the hermitian form is defined by
\begin{equation}
\spd uv=\spd uv_{\tilde{\mc H}_M}=\int_M(\spd {u_1}{v_1}_{\mc P}+\spd {u_2}{v_2}_{\mc P}).
\end{equation}

Let $\mc F_{\tup{ferm}}(\mc H)$ be the Fermi Fock space generated by $\tilde{\mc H}$ and let $a(u)$ \resp $a^*(u)$ be the corresponding annihilation \resp creation operators. These operators satisfy the anti-commutation rules
\begin{equation}\lae{6.4b}
\begin{aligned}
\{a(u),a(v)\}_+&=0\qq \A\, u,v\in \tilde{\mc H}_M\\
\{a^*(u),a^*(v)\}_+&=0\qq\A\, u,v\in \tilde{\mc H}_M
\end{aligned}
\end{equation}
and
\begin{equation}\lae{6.5b}
\{a(u),a^*(v)\}_+=\spd uv \id\qq\A\, u,v\in \tilde{\mc H}_M.
\end{equation}
A CAR relation is a triple $\{\mc H,\mc A(\mc H), b\}$, where $\mc H$ is a complex Hilbert space, $\mc A(\mc H)$ a unital $C^*$-algebra and $b$ a complex antilinear map from $\mc H$ to $\mc A(\mc H)$ such that its values satisfy the anti-commutation rules \re{6.4b} and \re {6.5b}.

We define $\mc H$ to be the Hilbert space generated by
\begin{equation}
\set{(\fv{Gu}M,\fv{D_\nu Gu}M)}{u\in C^\un_c(E,\hat E_2)}\su L^2(M,\hat M\times \hat M),
\end{equation}
where $\nu$ is the future directed normal to $M$,
and $b$ by
\begin{equation}
b(\fv{Gu}M)=a(\fv{Gu}M),
\end{equation}
where, by abusing the notation, we identified
\begin{equation}
\fv{Gu}M\equiv(\fv{Gu}M,\fv{D_\nu Gu}M).
\end{equation}
$\mc A(\mc H)$ is the $C^*$-algebra generated by the elements
\begin{equation}
\{b(\fv{Gu}M, b^*(\fv{Gu}M), \id\}\qq u\in C^\un_c(E,\hat E_2).
\end{equation}

The map
\begin{equation}
u\in C^\un_c(E,\hat E_2)\ra b(\fv{Gu}M),
\end{equation}
also denoted by $b$, such that
\begin{equation}
b(u)=b(\fv{Gu}M)
\end{equation}
is a distribution.
\bl
The map $b$ is an antilinear distribution in $C^\un_c(E,\hat E_2)$, or a distribution if we regard $C^\un_c(E,\hat E_2)$ as a real vector space. Moreover, $b$ is a weak solution of the fermionic Wheeler-DeWitt equation
\begin{equation}
Hb=0.
\end{equation}
\el
\bp
Let us first prove that $b$ is a distribution. We know that
\begin{equation}
\fv{Gu}M\q\wed\q \fv{D_\nu Gu}M
\end{equation}
have compact support, even uniformly compact support if $\supp u$ is contained in a compact $K\su E$. This statement is also valid for
\begin{equation}
\fv{G_+u}M\q\wed \q \fv{G_-u}M
\end{equation}
and the corresponding normal derivatives. $G_+u$ satisfies the hyperbolic equation
\begin{equation}
HG_+u=u
\end{equation}
with vanishing Cauchy conditions on a suitable Cauchy hypersurface, hence, for any compact subsets $K,K'\su E$ there exists an integer $m\in\N$ and a constant $c$ such that
\begin{equation}
\norm{\fv{G_+u}M}_{K'}+\norm{\fv{D_\nu G_+u}M}_{K'}\le c \abs u_{m,K}\qq\A\, u\in \mc D_K(E,\hat E_2),
\end{equation}
where $\mc D_K$ is the space of $u$'s with support in $K$.

The same result is also valid for $G_-u$ and hence for
\begin{equation}
Gu=G_+u-G_-u.
\end{equation}
This proves that $b$ is a distribution.

By the definition of weak derivatives of a distribution we have
\begin{equation}
\begin{aligned}
\spd{Hb}u=b(Hu)=b(\fv{GHu}M)=b(0)=0,
\end{aligned}
\end{equation}
since
\begin{equation}
GHu=0\qq\A\, u\in C^\un_c(E,\hat E_2).
\end{equation}

The corresponding quantum field $\Psi$ is defined by 
\begin{equation}
\Psi(u)=\frac1{\sqrt 2}(b(u)+b^*(u))\qq\A\,u\in C^\un_c(E,\hat E_2).
\end{equation}
$\Psi(u)$ is self-adjoint and a weak solution of the fermionic Wheeler-DeWitt equation. It satisfies the anti-commutation rules
\begin{equation}
\{\Psi(u),\Psi(v)\}_+=\rec \spd{\fv{Gu}M}{\fv{Gv}M}_M
\end{equation}
as one easily checks.

The full quantum field $\hat \F$ for the solutions of the Wheeler-DeWitt equation is defined by
\begin{equation}
\hat\F=(\F,\Psi).
\end{equation}
\ep

\section{The Haag-Kastler axioms}
Dimock generalized in \cite{dimock:qft} the Haag-Kastler axioms for local observables  in Minkowski space by considering local observables in a general globally hyperbolic spacetime. Dimock's ideas can also be applied in the present situation. We first look at the bosonic case.

\bd
Let $\eS\not=\Om\su E$ be an open relatively compact set and $M$ a Cauchy hypersurface in $E$, then we define $\mc A_M(\Om)$ to be the $C^*$-algebra generated by
\begin{equation}
\set{e^{i\F_M(u)}}{u\in C^\un_c(\Om,\hat E_1)}.
\end{equation}
We also define $\mc A_M$ to be the $C^*$-algebra generated by 
\begin{equation}
\set{e^{i\F_M(u)}}{u\in C^\un_c(E,\hat E_1)}.
\end{equation}
Finally, we define
\begin{equation}
\Lam=\set{\Om\su E}{\eS\not=\Om\;\wed\; \tup{$\Om$ open and relatively compact}}.
\end{equation}
\ed
\br\lar{6.2}
Let $M$, $M'$ be two Cauchy hypersurfaces in $E$ and 
\begin{equation}
T:\mc A_M\ra \mc A_{M'}
\end{equation}
the $^*$-isomorphism in \frr{5.7}, then
\begin{equation}
T(\mc A_M(\Om))=\mc A_{M'}(\Om)\qq\A\, \Om\in\Lam.
\end{equation}
\er
The collection 
\begin{equation}
\set{\mc A_M(\Om}{\Om\in\Lam}
\end{equation}
forms a \tit{net} of subalgebras of $\mc A_M$ as defined in \cite[Definition 2]{haag:net}.

Dimock considered these nets of local algebras in case when $E$ is a globally hyperbolic spacetime and listed five axioms satisfied by them. Four of the axioms are also valid in the present situation and will be described subsequently. The fifth, the so-called covariance axiom, is a bit more difficult to translate. The bundles and the operators are certainly covariant with respect to coordinate  and gauge transformations, but the covariance axiom postulates that an isometry of the underlying spacetime should induce an isomorphism of the local algebras. At the moment we do not know how to translate this axiom. 

We shall now list the four axioms for a fixed  Cauchy hypersurface $M\su E$.

\bn \tbf{Axiom 1} \; The family $\set{\mc A_M(\Om}{\Om\in\Lam}$ forms a net of local observables, i.e.,
\begin{equation}
\Om\su\Om'\im \mc A_\Om\su \mc A_{\Om'}
\end{equation}
and $\mc A_M$ is the closure of
\begin{equation}
\uuu_{\Om\in\Lam}\mc A_M(\Om).
\end{equation}
\en
This axiom is certainly satisfied as well as
\bn \tbf{Axiom 2} (Primitivity)\; $\mc A_M$ is primitive, i.e., it has a faithful irreducible representation.
\en

\bn \tbf{Axiom 3} (First causality)\; If $\Om$ is spacelike separated from $\Om'$, then
\begin{equation}
[\mc A_M(\Om),\mc A_M(\Om')]=0.
\end{equation}
Spacelike separated means that there is no causal curve joining a point in $\Om$ to a point in $\Om'$.
\en
This axiom is also satisfied since we deduce from  \re{5.11}, \fre{5.12}, \frt{5.6} and the properties of a Weyl system
\begin{equation}
W([u])W([v])=W([u]+[v])=W([v])W([u])
\end{equation}
for all $(u,v)\in C^\un_c(\Om,\hat E_1)\times C^\un_c(\Om',\hat E_1)$.

\bn \tbf{Axiom 4} (Second causality)\; If $\Om$ is causally dependent on $\Om'$, then
\begin{equation}
\mc A_M(\Om)\su \mc A_M(\Om').
\end{equation}
\en
$\Om$ is said to be causally dependent on $\Om'$ if there exists a Cauchy hypersurface $M'$ such that every endless causal curve through $p\in\Om$ intersects $M'\ii \Om'$. Hence, if $u\in C^\un_c(\Om,\hat E)$, then
\begin{equation}
\supp (Gu)\ii M'\su J^E(\supp u)\ii M'\su \Om'.
\end{equation}
From the arguments in \cite[p. 226]{dimock:qft} we then deduce that there exists $v\in C^\un_c(\Om')$ such that 
\begin{equation}
Gu=Gv
\end{equation}
and we conclude
\begin{equation}
\F_M(u)=\F_M(v)
\end{equation}
and therefore
\begin{equation}
\mc A_M(\Om)\su\mc A_M(\Om').
\end{equation}
Dimock only considered the Klein-Gordon operator but his arguments are valid for any self-adjoint normally hyperbolic operator.

In the fermionic case the four axioms are also valid, where in Axiom 3 the commutation brackets have to be replaced the anti-commutation braces. The proofs are almost identical. In case of Axiom $3$ we observe that
\begin{equation}
\{\Psi(u),\Psi(v)\}_+=\rec\spd{\fv{Gu}M}{\fv{Gv}+}_M=0,
\end{equation}
if $u\in C^\un_c(\Om,\hat E_2)$ and $v\in C^\un_c(\Om',\hat E_2)$, while in the other cases the proofs are literally identical.

\bibliographystyle{hamsplain}
\providecommand{\bysame}{\leavevmode\hbox to3em{\hrulefill}\thinspace}
\providecommand{\href}[2]{#2}



\end{document}